\documentclass[conference]{IEEEtran}
\IEEEoverridecommandlockouts
\usepackage{amsmath,amssymb,amsfonts}
\usepackage{algorithmic}
\usepackage{graphicx,color}
\usepackage{textcomp}
\usepackage{caption}
\usepackage{xcolor}
\usepackage{multirow}
\usepackage{svg}
\usepackage{hyperref}
\usepackage{float}
\usepackage{subcaption}
\usepackage[linesnumbered,ruled,vlined]{algorithm2e}
\usepackage{cite}
\def\BibTeX{{\rm B\kern-.05em{\sc i\kern-.025em b}\kern-.08em
    T\kern-.1667em\lower.7ex\hbox{E}\kern-.125emX}}
\begin{document}

\bstctlcite{IEEEexample:BSTcontrol}

\definecolor{red}{RGB}{255, 0, 0}
\definecolor{blue}{RGB}{0, 0, 255}

\title{MobRFFI: Non-cooperative Device Re-identification for Mobility Intelligence}

\author{
    \IEEEauthorblockN{Stepan Mazokha, Fanchen Bao, George Sklivanitis, Jason O. Hallstrom}
    \IEEEauthorblockA{\textit{I-SENSE (The Institute for Sensing And Embedded Network Systems Engineering)} \\
    \textit{Florida Atlantic University}\\
    Boca Raton, FL, USA \\
    \{smazokha2016, fbao2015, gsklivanitis, jhallstrom\}@fau.edu}
}

\maketitle

\begin{abstract}

    WiFi-based mobility monitoring in urban environments can provide valuable insights into pedestrian and vehicle movements. However, MAC address randomization introduces a significant obstacle in accurately estimating congestion levels and path trajectories. To this end, we consider radio frequency fingerprinting and re-identification for attributing WiFi traffic to emitting devices without the use of MAC addresses.

    We present MobRFFI, an AI-based device fingerprinting and re-identification framework for WiFi networks that leverages an encoder deep learning model to extract unique features based on WiFi chipset hardware impairments. It is entirely independent of frame type. When evaluated on the WiFi fingerprinting dataset WiSig, our approach achieves 94\% and 100\% device accuracy in multi-day and single-day re-identification scenarios, respectively.

    We also collect a novel dataset, MobRFFI, for granular multi-receiver WiFi device fingerprinting evaluation. Using the dataset, we demonstrate that the combination of fingerprints from multiple receivers boosts re-identification performance from 81\% to 100\% on a single-day scenario and from 41\% to 100\% on a multi-day scenario. 

\end{abstract}


\section{Introduction} \label{sec:introduction}

    Mobility monitoring in urban environments can provide valuable insights into pedestrian and vehicle movements. Today, local governments can automate the acquisition of such data using video surveillance. However, public disapproval of computer vision approaches due to privacy concerns opens opportunities for research into alternative, privacy-centric solutions, such as WiFi sensing. Modern mobile devices ubiquitously support the 802.11 standard and regularly emit WiFi probe requests for network discovery. We can passively monitor this traffic to estimate the levels of congestion in public spaces and localize devices to monitor motion trajectories.

    Previously, we deployed a testbed with 54 WiFi sensors along Clematis Street in downtown West Palm Beach and explored non-cooperative WiFi device localization. However, the presence of MAC address randomization \cite{martin2017study, vanhoef2016mac} limited our capacity to analyze device path trajectories over time and artificially inflated congestion estimates in public spaces. That is, rather than capturing a single \textit{physical} MAC address corresponding to a mobile device, the sensors detect multiple \textit{virtual} MAC addresses that are generated with each probe request emission. Since the occupancy counting and trajectory monitoring is based on the attribution of captured signals to device MAC addresses, randomization considerably complicates the analysis. To this end, prior work considers temporal pattern analysis and inspection of encapsulated fields, such as sequence numbers and preferred network lists \cite{uras2020wifi, tan2021efficient, uras2022mac, matte2016defeating}. However, the effectiveness of such methods can be hindered by simple obfuscation measures; varies across emitter devices and operating systems; and scales poorly with emitter count due to overlapping sequence numbers and other factors.

    To this end, preamble-based radio frequency fingerprinting and identification (RFFI) has emerged as an alternative to overcome MAC address randomization. Rather than using temporal characteristics or encapsulated data within probe requests, RFFI can produce unique device fingerprints (i.e., embeddings) based on hardware variations of the WiFi chipsets across any type of message frame.

    However, there are many challenges. For example, device fingerprints are subject to significant variability and deterioration in multi-day scenarios \cite{hanna2022wisig}. Fingerprint extraction models require extensive training, with inter-device discrimination suffering in tasks where new devices are continuously identified and added to the system (i.e., open-set scenario) \cite{shen2022towards}. Finally, fingerprints extracted by the model are associated with a specific receiver, which complicates device re-identification in large-scale environments.
    
    In this work, we demonstrate how state-of-the-art RFFI techniques can be used to perform robust WiFi device re-identification without using MAC addresses. We conduct experiments in an indoor environment, capturing randomly generated WiFi traffic across multiple sensors to evaluate our method. In particular, we provide the following contributions:

    % \textcolor{blue}{We argue that the majority of the associated re-identification challenges can be overcome either by re-framing system model requirements or by introducing model input, training, and framework improvements. For example, despite the device fingerprint deterioration in multi-day scenarios, RFFI has near-perfect performance in single-day scenarios. Even though extracted device fingerprints are receiver-specific and cannot be transferred, we can simultaneously capture and extract device fingerprints across multiple receivers by matching frame MAC addresses. Finally, the challenge of fingerprint discrimination can be improved by combining fingerprints from multiple receivers.}
    
    % In this paper, we consider these ideas and provide the following contributions:

    \begin{itemize}
    
        \item We introduce a WiFi device fingerprinting and re-identification framework. The system is built on an encoder-based deep learning architecture, which extracts hardware impairment features from WiFi preamble spectrograms. The result is a device fingerprint that can be used to match stored and new signal measurements using k-Nearest Neighbors (KNN) for re-identification.

        \item We explore the state-of-the-art WiSig dataset \cite{hanna2022wisig} and the corresponding WiFi device classification model as a benchmark for MobRFFI. Additionally, we collect a new 5.7 TB MobRFFI dataset for WiFi device fingerprinting with a focus on granular multi-receiver experimentation \cite{mazokha2024mobrffidataset}. 
        
        % \textcolor{blue}{This dataset provides one round of raw WiFi measurements across 4 days, spanning a month, which provides an opportunity to examine MobRFFI performance over extended periods of time. }

        % \textcolor{blue}{The dataset spans two days across 12 and 24 devices on day 1 and day 2, respectively. We perform 172 and 50 rounds of transmission across all devices on these dates, respectively, to evaluate single-day fingerprint stability.}
        
        \item We evaluate the performance of the MobRFFI framework across several modalities. First, we explore multiple signal pre-processing techniques and introduce an optimized WiFi spectrogram, boosting device classification accuracy by 13\%. We then exhaustively benchmark model performance on a closed-set scenario using WiSig, achieving 94\% classification accuracy using optimized channel-independent spectrograms.

        \item Finally, we examine the impact of combining fingerprints from multiple receivers on the fingerprint distances between emitting devices (i.e., fingerprint gap). Inter-fingerprint distance is key to evaluating the discriminative capacity of the re-identification system. Here, our results demonstrate a boost of known/unknown device classification accuracy from 81\% to 100\% on a single-day scenario and from 41\% to 100\% on a multi-day scenario. 
        
    \end{itemize}

    % {\bf Paper Organization.} Section \ref{sec:related_work} outlines recent advances in radio frequency fingerprinting and device re-identification. Section \ref{sec:system_design} describes the design and implementation of our re-identification framework. Section \ref{sec:datasets} briefly summarizes the state-of-the-art WiFi device fingerprinting dataset for multi-day method evaluation and describes the experimental setup we used to capture the MobRFFI dataset for granular multi-receiver experimentation. Section \ref{sec:evaluation} presents the experimental evaluation of our framework. Finally, Section \ref{sec:conclusions} describes findings and considers remaining challenges.

\section{Related Work} \label{sec:related_work}

    RFFI research has received significant attention over the last decade \cite{jagannath2022comprehensive, soltanieh2020review, xie2024radio}. Applications of this technique span device authentication, rogue device detection, identification and tracking of bad actors, and more. RFFI methods have proven successful for various RF mediums: ADS-B \cite{gritsenko2019finding, al2020exposing}, LR-WPAN \cite{merchant2018deep, peng2018design}, LoRa \cite{robyns2017physical, shen2022towards}, ZigBee \cite{peng2018design}, and most importantly for our work, WiFI \cite{brik2008wireless, gritsenko2019finding, hanna2022wisig, sankhe2019no, restuccia2019deepradioid, li2022radionet, jian2020deep}.

    We first review the most commonly applied and effective signal pre-processing techniques -- the first step in our device fingerprinting pipeline. Several approaches have proven effective when working with raw signal inputs (i.e., IQ samples). When dealing with low signal-to-noise (SNR) environments, such as crowded indoor and outdoor environments, the signal is passed through a filter pipeline, often including noise reduction, low-pass, and band-pass filters to improve the signal-to-noise ratio \cite{xie2024radio}. The signal is further processed through equalization to remove the impact of wireless channel variations \cite{sankhe2019no, restuccia2019deepradioid}. Importantly, the carrier frequency offset is not corrected at this stage, as it is correlated with hardware impairments of the transmitter \cite{jian2021radio, shen2022towards}. Finally, the signal is \textit{sliced}, which segments out fixed-length portions of the signal to generate inputs for the fingerprint extractor model. For WiFi, preamble-based signal slicing has proven effective \cite{soltanieh2020review}. 
    % \textcolor{blue}{In some methods, each slice is further processed via short-time Fourier transform (STFT) to produce spectrograms \cite{shen2021radio, shen2022towards, al2020exposing}, or used as input directly (i.e., in the time domain) \cite{merchant2018deep, hanna2020open}.}

    Next, we consider recent advancements in fingerprint extraction. Some hardware imperfections can be robustly inferred from signal characteristics, such as SYNC correlation, sampling frequency offset, carrier frequency offset, IQ errors, constellation trace figures, differential constellation trace figures, and more \cite{brik2008wireless, peng2018design}. Yet, their performance is limited due to ever-changing environmental conditions and the diversity of the fleet of observed devices. Fortunately, deep learning models (e.g., convolutional neural networks (CNNs)) have demonstrated near-perfect performance at extracting these features \cite{jian2020deep, sankhe2019no}. They are often adapted from architectures and loss functions developed for computer vision tasks, such as face recognition \cite{schroff2015facenet}. 
    % \textcolor{blue}{Some of the considered architectures include ResNet \cite{jian2021radio, al2020exposing} and long short-term memory (LSTM) networks for capturing temporal dependencies \cite{wang2020radio, ling2024rsbu}. }

    The models used vary based on scenarios. Closed-set circumstances are simpler and only require matching the newly obtained signal with one of the known devices. Multi-class classifiers with softmax probabilities have been used effectively for this task \cite{restuccia2019deepradioid, al2020exposing}. However, circumstances with large fleets of observed devices and open-set scenarios, where some of the devices aren't known to the network, are more challenging. Here, models can be used to extract and store device fingerprints for later matching via KNN or cosine similarity search \cite{gritsenko2019finding, soltanieh2020review}. 

    In our work, we consider the contributions described above and design a scalable fingerprinting pipeline that incorporates preamble-based signal slicing, applies short-time Fourier transform (STFT) to produce spectrogram images for each frame, extracts fingerprints using a CNN, and finally stores device embeddings in a vector database. We then implement a multi-receiver device matching process, yielding a robust device re-identification method.

\section{System Design} \label{sec:system_design}

    In this section, we present the design of MobRFFI, our framework for device fingerprinting and re-identification based on hardware impairment features.

    \subsection{Overview}

        \begin{figure*}[ht!]
            \centering
            \includegraphics[width=0.9\textwidth]{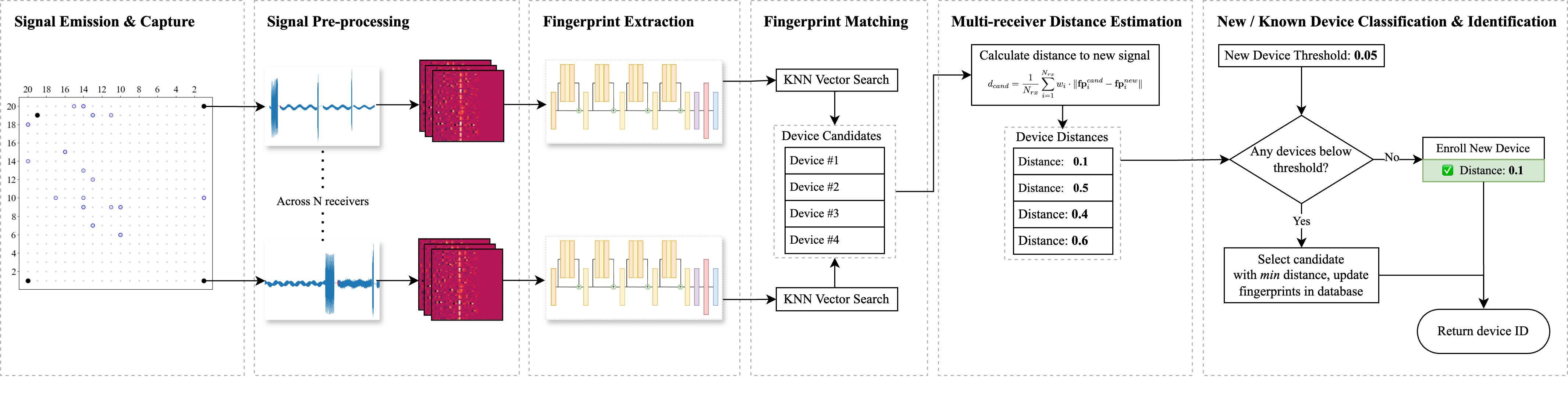}
            \caption{MobRFFI: device fingerprinting and re-identification framework.}
            \label{fig:rff_framework}
            % \Description{The high-level architecture of the MobRFFI framework for WiFi device fingerprinting and re-identification using hardware impairments.}
            \vspace{-2em}
        \end{figure*}

        The MobRFFI framework is illustrated in Fig. \ref{fig:rff_framework}. The system implements an encoder-based design, where a deep learning network is used to extract unique device fingerprints that characterize the hardware imperfections of a transmitting WiFi chipset. The framework consists of six stages. First, we capture raw WiFi IQ samples across several receivers. Next, we pre-process these samples to produce optimized channel-independent spectrograms. We then produce device fingerprints by passing these spectrograms through a pre-trained extractor model. The training process is illustrated in Fig. \ref{fig:model_training_pipeline} and is discussed later in this section. Following extraction, we perform fingerprint matching by finding a set of top-N nearest (previously captured) fingerprints in the database. Here, every matched fingerprint corresponds to one of the known WiFi devices in the network (i.e., already in the database). The matches are ranked using combined fingerprint distances across multiple receivers. Finally, pre-defined fingerprint distance thresholds are used to determine whether the new signal belongs to a \textit{known} device or a \textit{new} device. In the case of the latter, a new fingerprint is enrolled into the database.
    
    \subsection{Signal Capture and Pre-processing} \label{sec:system_design:preprocessing}

\begin{figure}[ht!]
            \centering
            \includegraphics[width=0.8\columnwidth]{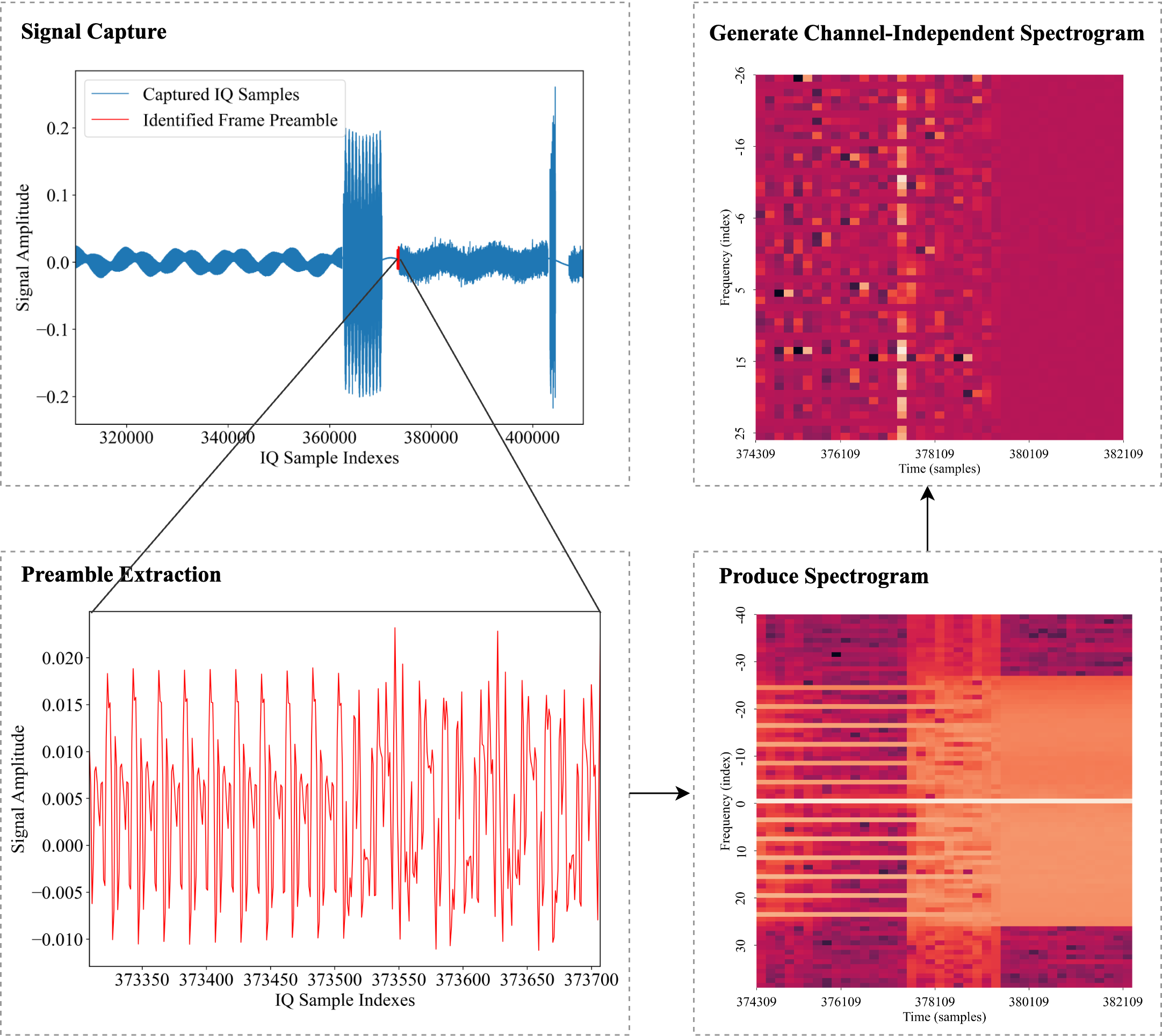}
            \caption{WiFi signal capture and pre-processing.}
            \label{fig:preprocessing_pipeline}
            % \Description{A high-level illustration of the MobRFFI WiFi signal pre-processing pipeline from raw IQ samples to optimized channel-independent spectrograms.}
            \vspace{-2em}
        \end{figure}

        The framework begins with signal capture and pre-processing, illustrated in Fig. \ref{fig:preprocessing_pipeline}. It consists of the following steps: signal capture, preamble extraction, spectrogram production, and the generation of a channel-independent spectrogram. The design of this pre-processing pipeline is informed by our comprehensive evaluation of various signal transformation techniques, detailed in Section \ref{sec:evaluation}.

        \subsubsection{Frame Detection, Slicing, and Filtering}
        
            % \textcolor{blue}{As discussed in Section \ref{sec:related_work}, there are multiple strategies for slicing raw signals into fixed-sized segments for fingerprinting. However, our work with 2.4 GHz WiFi protocol commands our choice of identifying WiFi frames and extracting their preambles. }
            
            According to the 802.11 standard, preambles are consistent across all types of frames. They consist of 10 Short Training Field (STF) sequences, a carrier prefix, and two Long Training Field (LTF) sequences. The STF and LTF sequences are used to estimate channel frequency offsets (CFO), channel state information (CSI), and received signal strength indicator (RSSI); equalize frames; and extract encoded data, such as MAC address and frame sequence number. Their consistency makes them well-suited for device fingerprinting.
    
            To identify the preambles, we avoid the simpler energy-based frame detection method \cite{hanna2022wisig} and leverage the \textit{WaveformAnalyzer} module from Matlab's open-source WLAN Toolkit implementation. We use it to perform coarse and fine cross-correlation between the received signal and the known STF and LTF patterns, respectively, to extract CFO and determine the start of each WiFi frame. Additionally, we use the toolkit to decode the frames and extract transmitter MAC addresses. 

            % \textcolor{blue}{Since we cannot guarantee perfect isolation of the testbed environment from rogue traffic, we use MAC addresses to discard frames that do not belong to our emitters. }
    
        \subsubsection{STFT Transformation and Subcarrier Removal}
        
            Next, we transform preambles into the frequency domain using STFT. We build on the methodology introduced by Shen et al. \cite{shen2022towards} and adapt it to the WiFi 802.11 protocol. 

            Consider a WiFi preamble $y = [y_1, y_2, ..., y_L]$, where $y_l$ is a complex IQ sample, and $L$ is the length of the preamble. The first step is to ensure a uniform power level across all samples via normalization (Eq. \ref{eq:preamble_normalization}). 

            \begin{equation} \label{eq:preamble_normalization}
                y^{\text{norm}} = \frac{y}{\sqrt{\frac{1}{L} \sum_{l=1}^{L} |y_l|^2}},
            \end{equation}

            Here, $|y_l|$ represents the magnitude of the $n$-th sample. Next, transform the normalized preamble $y^{\textbf{norm}}$ into a spectrogram using STFT (Eq. \ref{eq:stft}).
            
            \begin{equation} \label{eq:stft}
                S(k, m) = \sum_{n=0}^{N-1} y_n^{\text{norm}} w[n - mR] e^{-j 2\pi \frac{k}{N} n}
            \end{equation}

            Here, $k$ is a frequency bin index, $m$ is a time window index, $w[n]$ is a windowing function, $R$ represents the size of a hop between consecutive STFT windows in samples, and $N$ is the window size in samples. The resulting matrix represents both time-domain and frequency-domain characteristics of the WiFi preamble. The spectrogram may also be viewed as a 2D image, offering the potential to apply computer vision models for fingerprint extraction. 

            We set the STFT parameters according to our signal configuration. First, we address the number of available spectrogram windows. WiFi preambles are $20X$ shorter compared to LoRaWAN preambles (as used by \cite{shen2022towards}) and contain Orthogonal Frequency-Division Multiplexing (OFDM) STF and LTF sequences instead of LoRaWAN up-chirps. Therefore, we set the hop size $R = 0.1N$ to maximize the spectrogram temporal resolution. Second, we consider the size of the STFT window $N$, which must be sufficient to balance the temporal and frequency resolution of the spectrogram. Given the sampling rate of WiSig and MobRFFI capture, $f_s = 25 Msps$, and the WiFi subcarrier spacing $\Delta f = 312.5 kHz$, we calculate $N = \frac{f_s}{\Delta f} = 80$.

            Following this transformation, we construct a channel-independent spectrogram $Q(k, m)$ (Eq. \ref{eq:ch_ind}), as proposed by Shen et al. \cite{shen2022towards}. 

            \begin{equation} \label{eq:ch_ind}
                Q(k, m) = \frac{S(k, m+1)}{S(k, m)}
            \end{equation}

            Here, adjacent STFT windows are divided to suppress wireless channel effects while preserving device-specific characteristics of the captured signal.

            Next, we convert the channel-independent spectrogram $Q(k, m)$ to the decibel scale to emphasize power differences (Eq. \ref{eq:ch_ind_mag}), and standardize the spectrogram to normalize its mean $\mu$ and standard deviation $\sigma$ (Eq. \ref{eq:ch_ind_mag_std}).

            \begin{equation} \label{eq:ch_ind_mag}
                Q(k, m) = log_{\textbf{10}}(|Q(k, m)|^2)
            \end{equation}

            \begin{equation} \label{eq:ch_ind_mag_std}
                Q'(k, m) = \frac{Q(k, m) - \mu}{\sigma}
            \end{equation}
            
            Finally, we consider the removal of the guard and DC subcarriers. These frequencies are reserved to prevent interference between adjacent WiFi channels and do not carry any data. Since their contents do not represent information transmitted by the emitter, their removal can simplify the training process and further reduce the impact of noise on model performance. This step and its impact on device classification accuracy are evaluated in Section \ref{subsec:spec_optimization}.

    \subsection{Fingerprint Extraction}
    
        \subsubsection{Model Architecture} 
        
\begin{figure*}[ht!]
                \centering
                \includegraphics[width=0.7\textwidth]{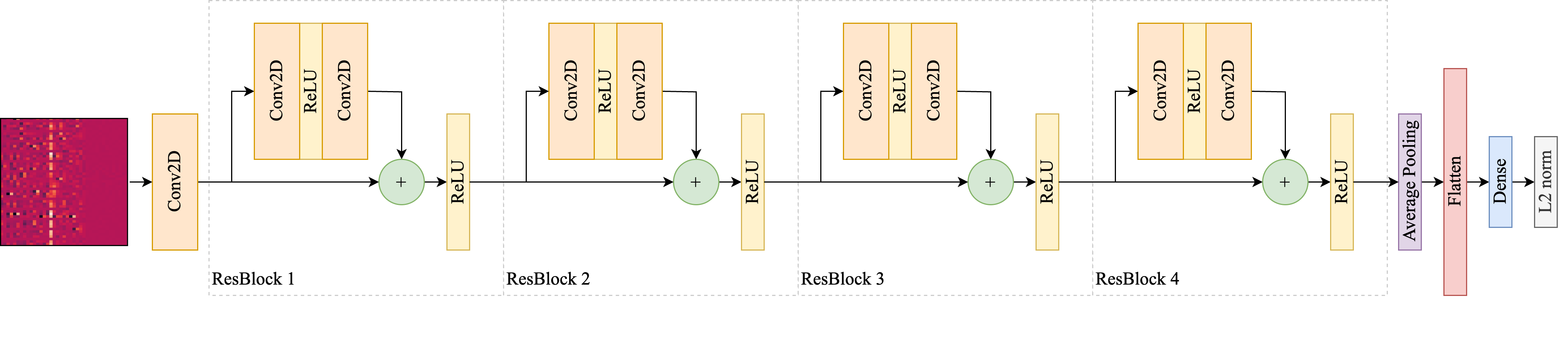}
                \caption{Fingerprint extractor model architecture.}
                \label{fig:model_architecture}
                % \Description{A high-level illustration of the architecture of the WiFi fingerprint extractor deep learning model.}
                \vspace{-2em}
            \end{figure*}

            We adopt a convolutional neural network design \cite{shen2022towards} based on ResNet \cite{he2016deep} to produce device fingerprints. The model layers are illustrated in Figure \ref{fig:model_architecture}. It begins with a convolutional layer, which applies learnable filters (kernels) to identify unique low-level features in spectrogram images, such as frequency shifts, hardware-specific anomalies, and more. Next, the model contains four residual blocks, which allow learning more complex, abstract features of signal impairments. As proposed by He et al. \cite{he2016deep}, each residual block has a skip link, which allows the skip of any layers that degrade the model performance and, therefore, avoids issues of expanding and vanishing gradients. The ResNet blocks are followed by the average pooling layer, which reduces the spatial dimensions of the feature maps from previous layers and keeps the most prominent features in the spectrogram. Its output is flattened into a 1-dimensional vector and passed through a fully connected dense layer to produce a more compact embedding. Finally, the L2 norm layer normalizes the output embeddings and focuses the model on extracting unique features of each device chipset without being influenced by the variations in signal strength. 
            % \textcolor{blue}{Overall, the model consists of 1,350,144 parameters; it is considered lightweight.}

        \subsubsection{Model Training}
        
\begin{figure}[ht]
                \centering
                \includegraphics[width=\columnwidth]{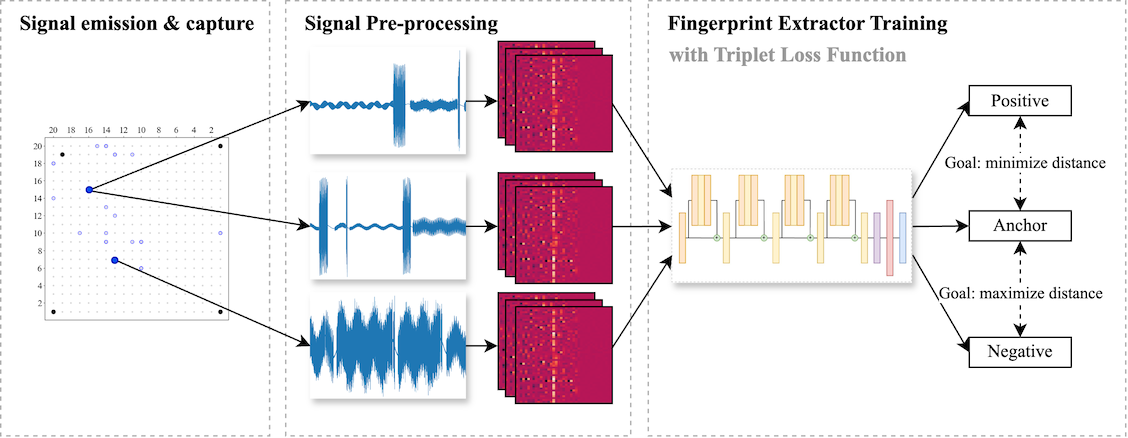}
                \caption{Fingerprint extractor training with triplet loss function.}
                \label{fig:model_training_pipeline}
                % \Description{Illustration of the WiFi fingerprint extractor model training process with a triplet loss function.}
                \vspace{-2em}
            \end{figure}

            The model training process is illustrated in Fig. \ref{fig:model_training_pipeline}. Similar to the main device fingerprinting pipeline, the process starts with capturing the signal, extracting preambles, and pre-processing them to obtain the optimized channel-independent spectrograms. These spectrograms are used as input to the deep learning model, which is trained to extract hardware impairment features unique to each device. 
            
            Model training is then performed using the triplet loss function, shown in Eq. \ref{formula:triplet_loss} \cite{schroff2015facenet}. 
            
            \begin{equation}
                L_{\text{triplet}} = \sum_{i,j,k}^{N} \left[ \| f(x_i) - f(x_j) \|_2^2 - \| f(x_i) - f(x_k) \|_2^2 + \alpha_{\text{triplet}} \right]_+
                \label{formula:triplet_loss}
            \end{equation}

            Here, $f(x_i), f(x_j), and f(x_k)$ represent the \textit{anchor}, \textit{positive} and \textit{negative} extracted feature vectors respectively. They are produced using randomly selected combinations of inputs from the training dataset and form the anchor-positive and anchor-negative pairs $\| f(x_i) - f(x_j) \|$ and $\| f(x_i) - f(x_k) \|$, respectively. However, while the anchor and the positive vectors are produced using signals from the same device, the negatives represent signals from a different transmitter. The $\alpha_{triplet}$ constant is a threshold margin enforced between these pairs. 
            % \textcolor{blue}{We select its value by evaluating model performance on a closed-set device evaluation scenario in the range between $0.1$ and $2.0$ with a step of $0.05$ and achieve the best results with $\alpha_{triplet} = 1.1$. }
            The purpose of the loss function is to minimize the distance between the anchor and a positive pair while maximizing the distance between the anchor and the negative pair.  

            During training, we optimize CNN weights using the \textit{Keras RMSprop} optimizer, with a learning rate of $0.001$. We also configure the training process to automatically reduce the learning rate and stop the training after the loss function does not decrease after 10 epochs.

    \subsection{Continuous Device Re-identification}

        After fingerprints are produced, our task is to resolve two classification problems. First, we must determine whether a new signal is emitted by a known device. If yes, the second task is to identify the known device in the database.

        To resolve these tasks, we implement a re-identification algorithm. We start with a list of device preambles across all receivers. First, we transform each preamble into optimized channel-independent spectrograms following our pipeline from Section \ref{sec:system_design:preprocessing} and produce device fingerprints. We obtain average extracted fingerprint embeddings $rx\_fps_i$, and RSSI values $rx\_rssi_i$ across all available frames within each receiver. We then retrieve the top N most relevant \textit{device candidates}. For this step, we use an open-source vector database, Chroma \cite{chromadb2024}. It implements a KNN-based vector search with cosine similarity as a distance metric to support vector retrieval. We perform this search and compile a list of the top-$K$ potential device matches across all receivers $RX$, where $K$ is determined experimentally. Next, we use Euclidean distance to measure the similarity of the input signal with a fingerprint from a corresponding receiver for each device candidate. To improve fingerprint discrimination, we combine fingerprint distances from multiple receivers into $D_c$ using RSSI values $rx\_rssi_i$ as weights. We produce weights by normalizing RSSI values from each receiver between $-100\ dBm$ and $0\ dBm$. Finally, we determine whether the newly extracted fingerprints correspond to one of the known devices by comparing the smallest combined distance with a pre-defined $threshold$. The latter is determined experimentally and is further discussed in Section \ref{sec:multirx_reid}. If such a device candidate exists, we associate the new signal with a known device. Otherwise, we enroll a new device into the database.

\section{Datasets} \label{sec:datasets}

    In this section, we provide an overview of the two datasets used to evaluate our device re-identification framework. First, we briefly review the state-of-the-art WiFi dataset WiSig \cite{hanna2022wisig}, designed for multi-day device fingerprinting. We describe its shortcomings and the motivation behind capturing our own dataset, which is focused on granular, multi-receiver features. For the latter, we describe the corresponding experimental setup and features.

    \subsection{WiSig Dataset}

\begin{figure}[ht!]
            \centering
            \begin{subfigure}[b]{0.4\linewidth}
                \includegraphics[width=\linewidth]{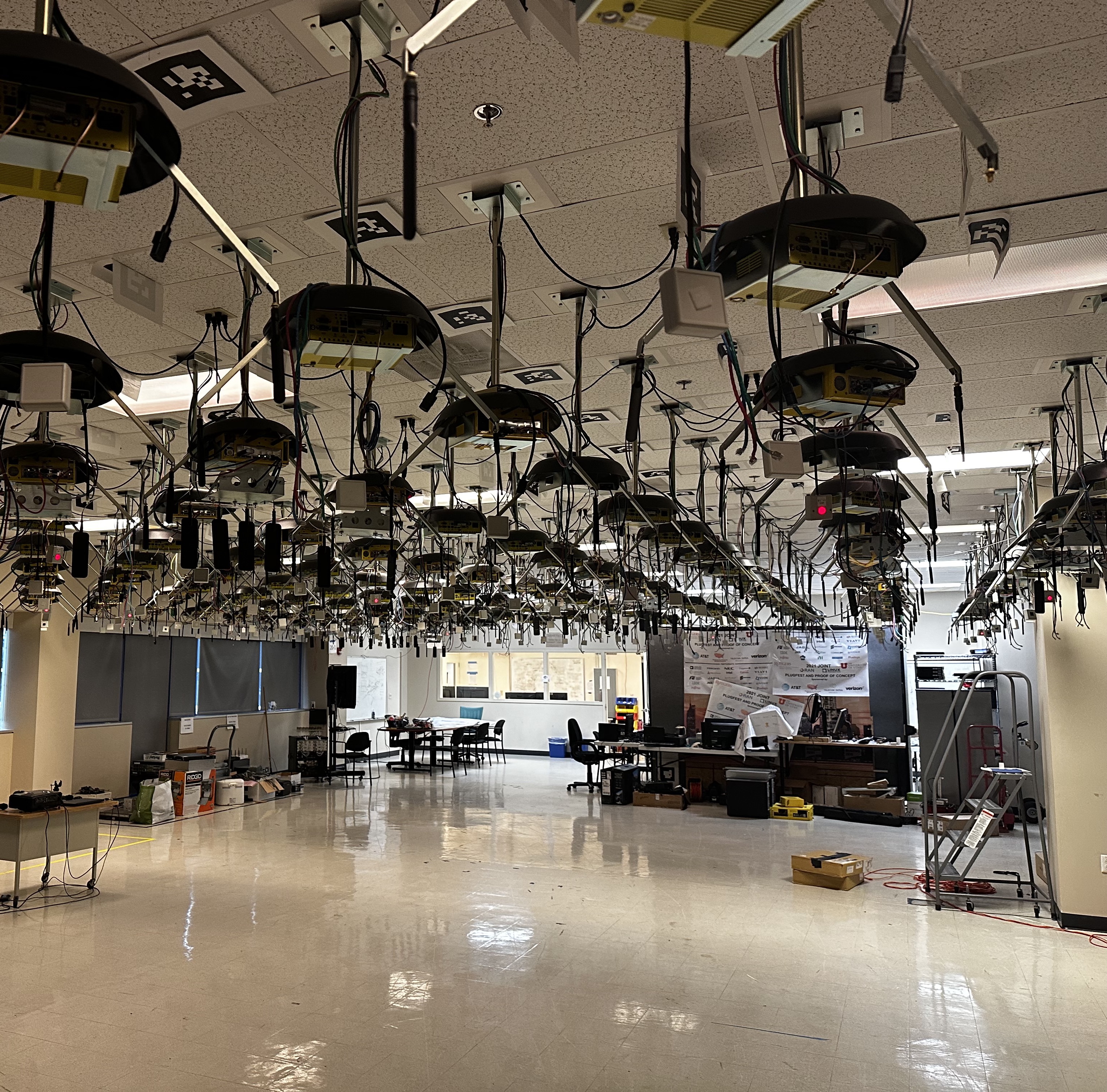}
                \caption{Orbit ceiling nodes.} \label{fig:exp_setup_facility}
            \end{subfigure}
            \hfill
            \begin{subfigure}[b]{0.4\linewidth}
                \includegraphics[width=\linewidth]{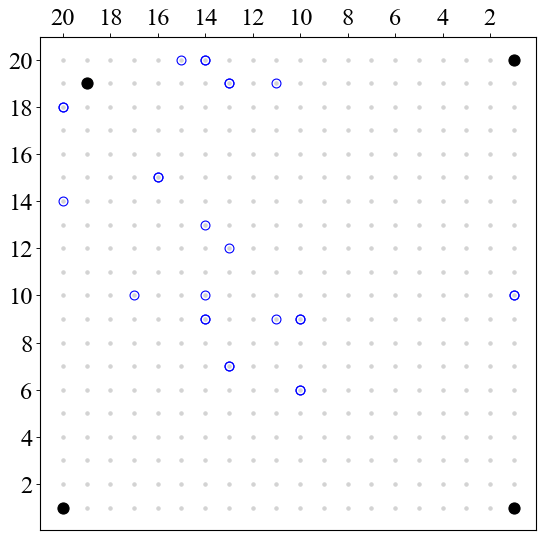}
                \caption{Node topology.} \label{fig:exp_setup_topology}
            \end{subfigure}
            
            \vspace{1em}
            \begin{subfigure}[b]{\linewidth}
                \centering
                \includegraphics[width=0.8\linewidth]{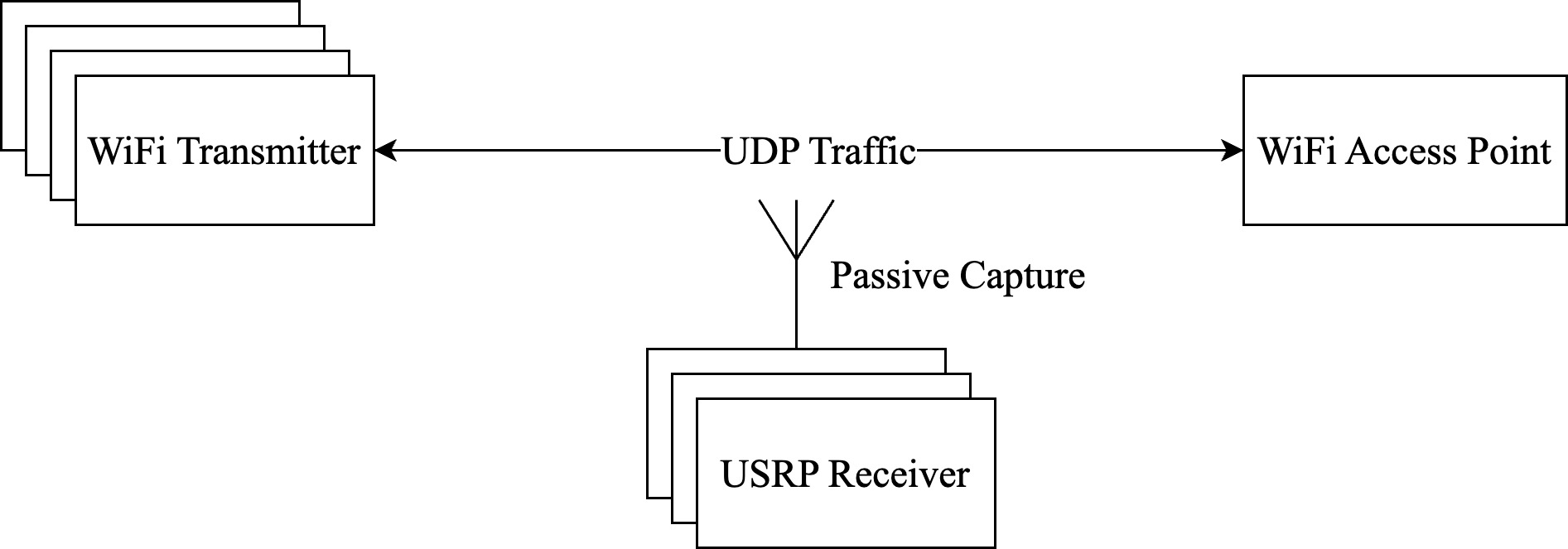}
                \caption{Traffic capture setup.} \label{fig:exp_setup_capture}
            \end{subfigure}
            \caption{Orbit experimental setup.} \label{fig:orbit_setup}
            % \Description{The illustrations of the Orbit testbed facility, emitter and receiver node topology, and the traffic capture setup.}
            \vspace{-2em}
        \end{figure}

        The WiSig dataset is a state-of-the-art WiFi device fingerprinting dataset captured in the Orbit testbed facility at Rutgers University in 2021 \cite{hanna2022wisig, raychaudhuri2005overview}. The testbed provides access to various models of ceiling-mounted emitters supporting WiFi transmission and several Universal Software Radio Peripheral (USRP) stations with antennas that support capturing signals in multiple bands, including 2.4 GHz, 5 GHz, and mmWave. A photo of the facility is illustrated in Fig. \ref{fig:exp_setup_facility}.

        % \textcolor{blue}{The dataset comprises 4 days of signal capture spanning one month and containing transmissions across 174 transmitting and 41 receiving devices. The number of devices varies across days, depending on node availability in the testbed. Each capture file contains raw IQ samples of 0.5 seconds of WiFi traffic. The researchers configured data transmission on a 2.4 GHz channel with a 25 Msps sampling rate. Each transmission involves three types of devices: WiFi transmitters (TX), a WiFi access point, and USRP receivers (RX), as illustrated in Fig. \ref{fig:exp_setup_capture}. Here, the UDP packets containing randomly generated payload data are transferred between a transmitter and an access point. Simultaneously, the USRP receivers capture the signal on the same WiFi channel. Only one emitter is active at a time, and only one 0.5-second signal capture between any transmitter/receiver pair is obtained during a single day.}
        
        We adopt the WiSig dataset as the initial benchmark for evaluating our device re-identification framework. In particular, we leverage multi-day signal capture data to evaluate model performance over extended periods of time. However, the signal from a single transmitter isn't captured across multiple receivers simultaneously, which limits our capacity to combine fingerprints from a single receiver. Further, the dataset provides no repeated signal captures during a 24-hour period, which we must explore to ensure that device re-identification is feasible during prolonged periods of time within a single day. This motivates our collection of the MobRFFI dataset.

    \subsection{MobRFFI Dataset}

        To address the shortcomings of the WiSig dataset, we collected the MobRFFI dataset \cite{mazokha2024mobrffidataset}. Its characteristics are shown in Table \ref{tab:mobrffi_dataset_stats}. Here, our priority is to evaluate the performance of our device re-identification framework on a scale of a single day while combining data from multiple receiving nodes. 

        \begin{table}[ht!]
            \small
            \centering
            \begin{tabular}{p{0.3\linewidth}|p{0.2\linewidth}|p{0.2\linewidth}}
            & \textbf{07/19/2024} & \textbf{08/08/2024} \\
            \hline
            Receivers & 4 & 3 \\
            \hline
            Training Rounds & 1 & 1 \\
            \hline
            Evaluation Rounds & 171 & 50 \\
            \hline
            Training Emitters & 29 & 28 \\
            \hline
            Testing Emitters & 12 & 24 \\
            \hline
            TX/RX Capture Files & 8,372 & 3,696 \\
            \hline
            Total Size, TB & 3 TB & 2.7 TB \\
            \end{tabular}
            \caption{MobRFFI dataset characteristics.}
            \label{tab:mobrffi_dataset_stats}
        \end{table}

        To ensure consistency between dataset experiments and evaluation, we reproduce the WiSig signal transmission setup in our dataset with a few modifications. First, we extend the capture window to 2 seconds per transmitter/receiver pair to ensure that we capture sufficient frames for experimentation. Second, we use disjoint sets of devices for training and testing the fingerprint extractor model, named \textit{seen} for the former and \textit{unseen} for the latter. A single round of signal capture is performed for the \textit{seen} set, while the signals from the \textit{unseen} set are performed repeatedly for a duration of 35 hours on Day 1 (\textit{July 19, 2024}), with 10-minute gaps, and 4 hours on Day 2 (\textit{August 8, 2024}), with 20-minute gaps. To achieve the time gaps between rounds, we consider the time required for a single transmitter/receiver capture pair (e.g., 30 seconds) and reduce the size of the \textit{unseen} set to 12 and 24 devices for the two days, respectively.

\section{Experimental Evaluation} \label{sec:evaluation}

    In this section, we discuss our design decisions and evaluate the performance of MobRFFI. First, we consider the effectiveness of preamble-based signal slicing and its impact on device classification accuracy. Second, we evaluate the impact of removing guard and DC subcarriers on the performance of the fingerprint extractor model. Third, we explore the stability of device fingerprints over the span of a day. 
    % and examine the multi-day volatility of CFO values in both the WiSig and MobRFFI datasets. 
    Next, we evaluate signal pre-processing techniques and RFFI model architectures in a closed-set scenario. Finally, we proceed to open-set evaluation. Here, we demonstrate the impact of combining fingerprints from multiple receivers on the fingerprint distance between emitting devices and RFFI performance on an open-set seen/unseen device classification problem.

    \subsection{Spectrogram Optimization} \label{subsec:spec_optimization}

\begin{figure}[ht!]
            \centering
            \begin{subfigure}{0.45\columnwidth}
                \centering
                \includegraphics[width=\linewidth]{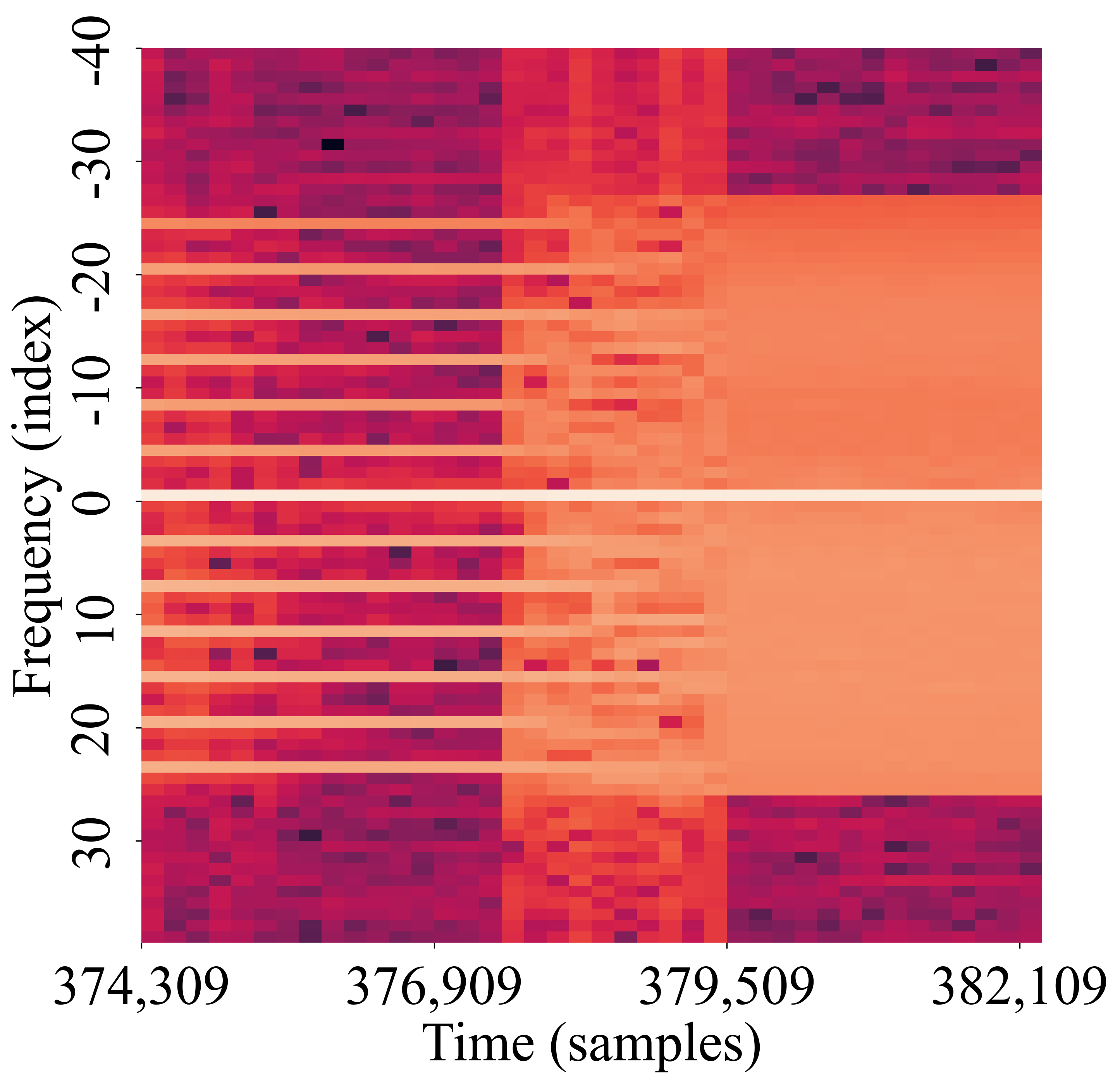}
                \caption{Spectrogram with all subcarriers (device classification accuracy: 81\%).}
                \label{fig:spec_design_full}
            \end{subfigure}
            \hspace{0.05\columnwidth}
            \begin{subfigure}{0.45\columnwidth}
                \centering
                \includegraphics[width=\linewidth]{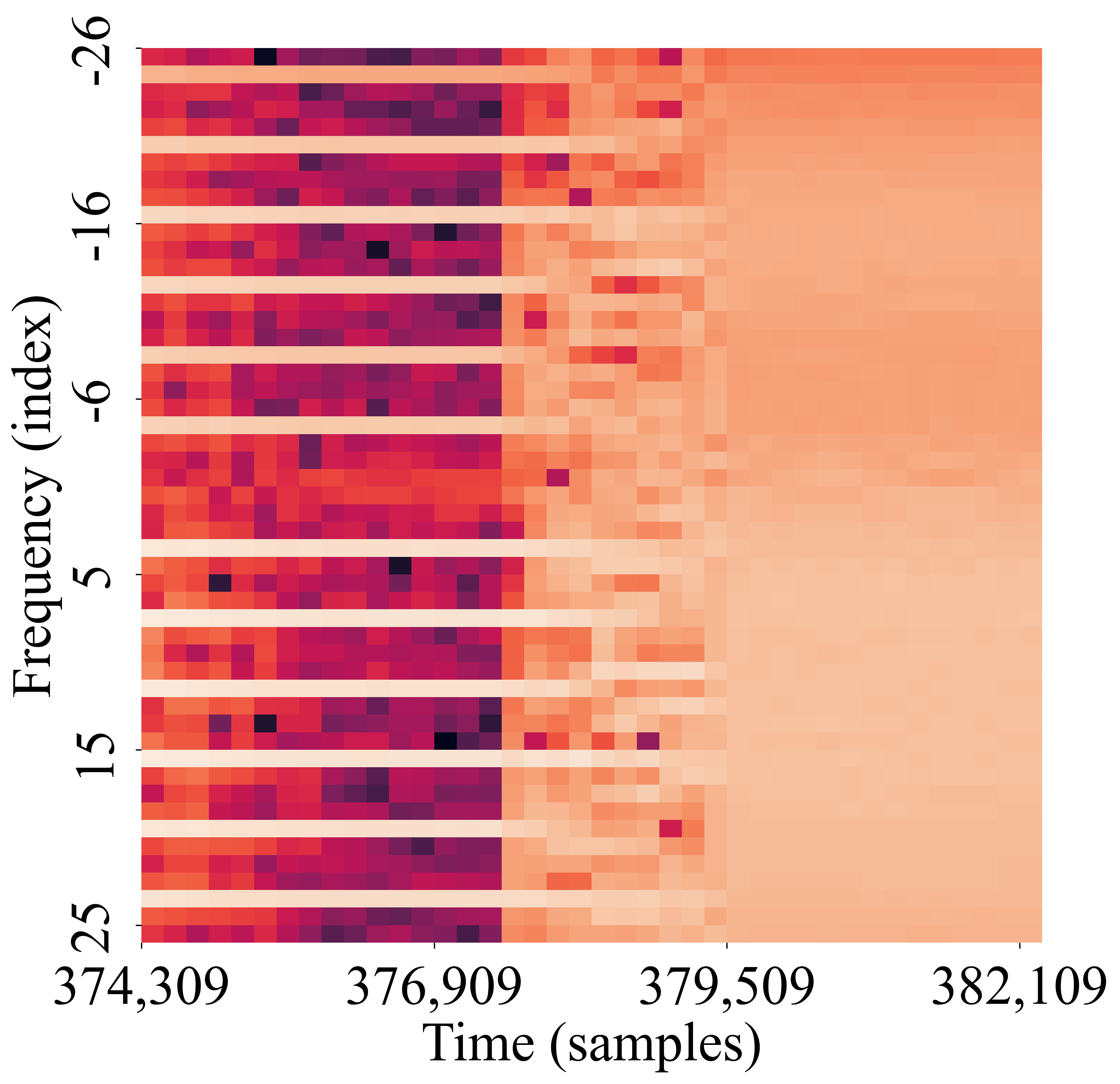}
                \caption{Spectrogram without guard subcarriers (device classification accuracy: 94\%).}
                \label{fig:spec_design_no_guards}
            \end{subfigure}
            \caption{Spectrogram optimization.}
            \label{fig:spec_design}
            % \Description{Illustration of the WiFi preamble spectrograms before and after removing guard and DC subcarriers.}
            \vspace{-1.4em}
        \end{figure}

        In Section \ref{sec:system_design:preprocessing}, we noted that guard and DC subcarriers do not contribute to the quality of device fingerprints since they do not carry any data. Therefore, we remove the DC and guard subcarriers; we evaluate the performance of both full and reduced-size spectrograms on our best-performing model in a closed-set device classification scenario to validate our decision. We illustrate the spectrogram heatmaps in Fig. \ref{fig:spec_design_full} and Fig. \ref{fig:spec_design_no_guards}, respectively. Here, the horizontal axis represents indices of the STFT windows, while the vertical axis represents subcarrier indices. The intensity of each tile corresponds to the power at a specific time-frequency point. The full-spectrogram model results in 81\% classification accuracy. However, its optimized counterpart increases accuracy to 94\%, giving it a 13\% boost in performance. Therefore, we use reduced-size spectrograms to perform the remaining experiments.

    \subsection{Single-day Fingerprint Temporal Stability}

        The long-term stability of device fingerprints is widely claimed to be one of the major shortcomings of RFFI methods \cite{hanna2022wisig}. However, for the purposes of device re-identification to enable trajectory tracking in streetscape environments, single-day fingerprint consistency is sufficient.
        
\begin{figure}
            \centering
            \includegraphics[width=0.6\linewidth]{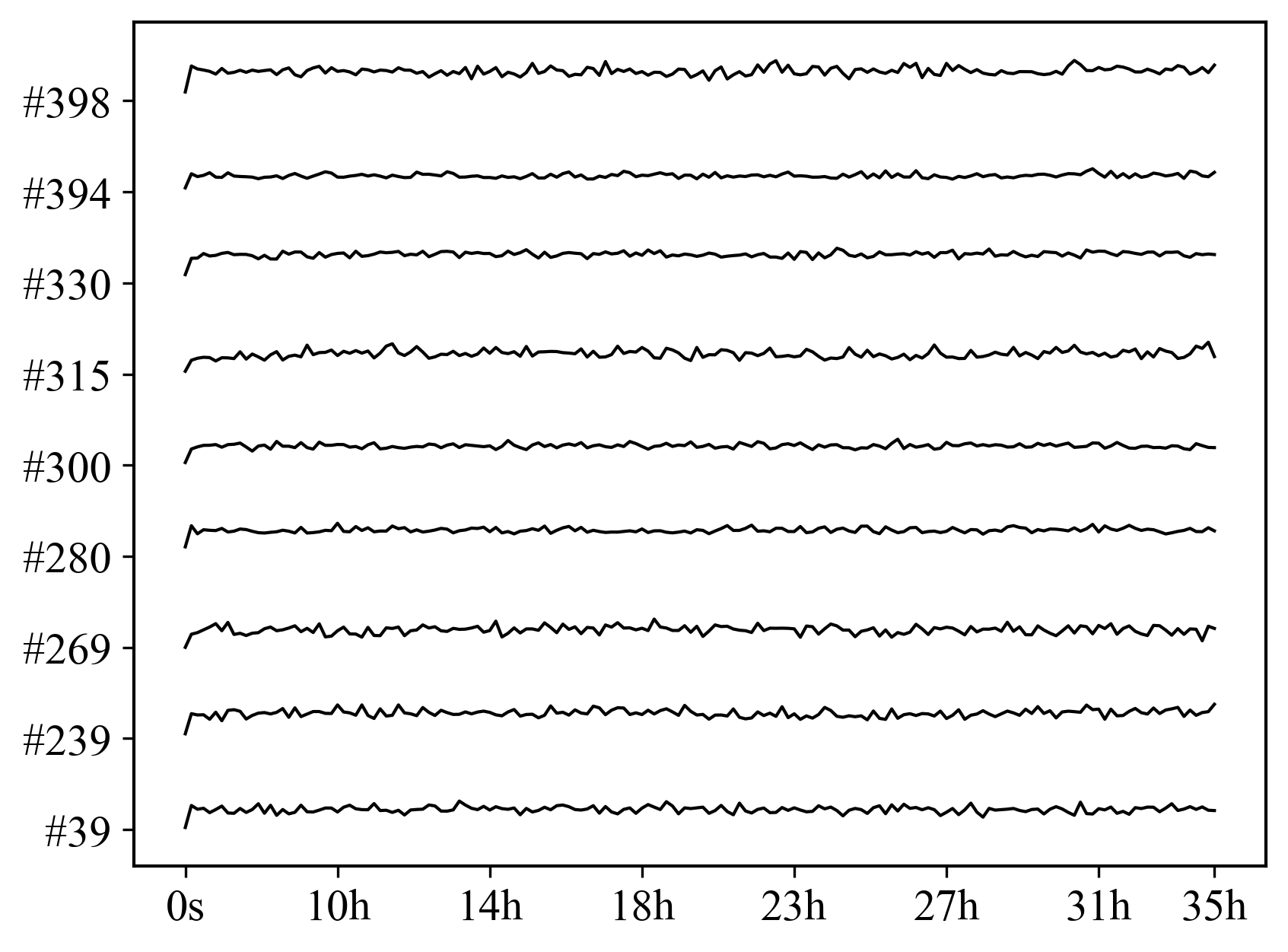}
            \caption{Fingerprint temporal stability, \textit{8/8/2024} dataset.}
            \label{fig:temporal_stability_evaluation}
            % \Description{Illustration of WiFi device fingerprint stability throughout the period of a single day.}
        \end{figure}

        We evaluate fingerprint stability using the MobRFFI dataset captured on \textit{July 19, 2024}. Taking the first captured round of fingerprints as a reference, we estimate their Euclidean distance to each of the subsequent 171 fingerprinting rounds, captured at 20-minute intervals. The normalized distances are illustrated in Figure \ref{fig:temporal_stability_evaluation}. Here, the horizontal axis represents the index of the signal capture round, and the vertical axis represents the IDs of the transmitting modes. Fingerprint distances are averaged across 100 frames in each round. 

        % TODO: 3% is a value from memory; update when CA-AI server is back online...
        
        From this figure, we can confirm that the fingerprints remain stable during the observed time span with a standard deviation not exceeding 3\%. Such a time window is sufficient to oversee device movement activity throughout the area visible to WiFi sensors, even if the fingerprints demonstrate significant degradation over days or weeks \cite{hanna2022wisig}. 

    \subsection{Closed-set Device Classification}

        We begin our evaluation of the fingerprint extractor model on a closed-set scenario, where the same group of devices is first enrolled and later re-identified using a newly captured signal. Here, we consider the choices of the signal pre-processing techniques and the loss function discussed in previous sections.

        \subsubsection{Evaluation of Pre-processing Techniques and RFFI Model Architectures}

            \begin{table*}[ht!]
                \centering
                \footnotesize
                \caption{Evaluation of signal pre-processing techniques and model architectures in a closed-set scenario on WiSig.}
                \begin{tabular}{l|ccc|ccc}
                    \hline
                    \multicolumn{7}{c}{\textbf{Fingerprint Extractor} (Tested on Day 4)} \\
                    \hline
                    & \multicolumn{3}{c|}{Seen Devices} & \multicolumn{3}{c}{Unseen Devices} \\
                    Training Set & Day 1 & Day 1+2 & Day 1+2+3 & Day 1 & Day 1+2 & Day 1+2+3 \\
                    \hline
                    \multicolumn{7}{l}{\textbf{Same-day}} \\
                    Raw & 100.00\% & 100.00\% & 100.00\% & 100.00\% & 100.00\% & 100.00\% \\
                    Equalized & 99.00\% & 100.00\% & 100.00\% & 100.00\% & 100.00\% & 100.00\% \\
                    Ch.-ind. & 100.00\% & 100.00\% & 100.00\% & 100.00\% & 100.00\% & 100.00\% \\
                    Eq. + Ch.-ind. & 100.00\% & 100.00\% & 99.00\% & 99.00\% & 99.00\% & 98.00\% \\
                    \hline
                    \multicolumn{7}{l}{\textbf{Multi-day}} \\
                    Raw & 67.00\% & 100.00\% & 100.00\% & 70.00\% & 39.00\% & 37.50\% \\
                    Equalized & 54.00\% & 79.00\% & 88.00\% & 59.00\% & 55.00\% & 66.00\% \\
                    Ch.-ind. & 78.00\% & 90.50\% & 99.00\% & \textbf{94.00\%} & 75.00\% & 69.83\% \\
                    Eq. + Ch.-ind. & 76.00\% & 84.00\% & 85.00\% & 70.00\% & 61.00\% & 77.00\% \\
                    \hline
                    \hline
                    \multicolumn{7}{c}{\textbf{WiSig Device Classifier} (Tested on Day 4)} \\
                    \hline
                    & \multicolumn{3}{c|}{Seen Devices} & \multicolumn{3}{c}{Unseen Devices} \\
                    Training Set & Day 1 & Day 1+2 & Day 1+2+3 & Day 1 & Day 1+2 & Day 1+2+3 \\
                    \hline
                    \multicolumn{7}{l}{\textbf{Same-day}} \\
                    Raw & 99.96\% & 100.00\% & 100.00\% & N/A & N/A & N/A \\
                    Equalized & 100.00\% & 100.00\% & 100.00\% & N/A & N/A & N/A \\
                    \hline
                    \multicolumn{7}{l}{\textbf{Multi-day}} \\
                    Raw & 58.08\% & 80.42\% & 73.29\% & N/A & N/A & N/A \\
                    Equalized & 60.50\% & 83.21\% & 83.54\% & N/A & N/A & N/A \\
                    \hline
                \end{tabular}
                \label{tab:closed_set_wisig_comparison}
            \end{table*}

            First, we compare our extractor performance to the WiSig device classification model on the WiSig dataset. We perform our evaluation across a wide range of parameters. First, we consider the number of days of data used to train both models. Second, we evaluate the signal pre-processing techniques used to produce input spectrograms. Here, we compare the results for raw signals, equalized frames, channel-independent spectrograms, and combined equalized and channel-independent spectrograms. Third, we consider the performance of both models on a subset of devices used for training the model (i.e., seen devices), as well as those that have not been seen before (i.e., unseen devices). For this experiment, we find a set of 25 devices for which we can consistently extract and pre-process 500 frames across each of the 4 days of signal capture. To train the model, we use 500 frames from 19 devices and augment the training dataset by replication and the addition of Gaussian noise. During the evaluation, we enrolled devices using 50 frames from day 1 for each device and evaluated classification performance using 100 frames from day 4 to match the evaluation provided in the WiSig paper \cite{hanna2022wisig}. A single receiver signal was used for this experiment.

            The results of our evaluation are summarized in Table \ref{tab:closed_set_wisig_comparison}. Since the WiSig device classifier design does not allow for the evaluation of model performance on an unseen set of emitters, we mark the corresponding tiles as N/A. We quickly see that both models demonstrate near-perfect performance in a same-day scenario, where devices are re-identified using a signal from day 1. In a multi-day scenario for a seen set of devices, as originally presented by Hanna et al. \cite{hanna2022wisig}, the WiSig model demonstrates improved performance on equalized frames, with training data combined from two and three days. The performance of the extractor model requires a closer look. The model achieves 100\% and 88\% classification accuracy on the raw and equalized datasets, respectively. However, it fails to reach acceptable performance on an unseen set of devices, which is a key metric for its usability in a real-world scenario. The performance is slightly improved when channel-independent spectrograms are obtained from equalized frames. The best performance of 94\% is achieved by channel-independent spectrograms without equalization when trained on a single-day dataset.
        
        \subsubsection{Evaluation of Single-receiver and Multi-receiver Device Classification for MobRFFI}

            It is important to note that the results of both models are heavily dependent on the amount of captured signal, the SNR, the line-of-sight conditions, and the distance between transmitting and receiving devices. 
            
            To demonstrate this, we evaluate the extractor model performance on the MobRFFI dataset. The results are summarized in Table \ref{tab:closed_set_mobrffi_comparison}. The model training dataset consisted of signals from 19 devices, but with only 200 frames per device available for training, caused by lower SNR during the data collection, which reduced the capacity to identify frames in raw captured IQ samples. We used 50 frames from 6 devices for enrollment and 100 frames for identification. 
            
            While both WiSig and MobRFFI datasets were captured in the same facility, different sets of emitting and receiving sensors were used for signal capture. Additionally, the single-receiver classification was performed using the signal from $node1-1$, located in the distant corner of the testbed environment with a large number of visual obstacles from the receivers. As a result, while the model demonstrates 80\% single-day single-receiver performance, it delivers only 40\% performance in a multi-day scenario.
            
            We address this situation by combining fingerprints from multiple receivers captured simultaneously and weighing fingerprint distances using the normalized RSSI values. The latter can provide a path to reduce the impact of low-quality fingerprints on device classification in dynamically changing environments.

            With RSSI-based weighting used to combine fingerprints from three receivers, the model classification performance is significantly improved from 81\% to 100\%, and 42\% to 100\% in single-day and multi-day scenarios, respectively.

            \begin{table}[ht!]
                \centering
                \footnotesize
                \caption{Evaluation of single-receiver and multi-receiver device classification in a closed-set scenario on MobRFFI dataset.}
                \begin{tabular}{l|ccc}
                    \hline
                    & \multicolumn{3}{c}{Unseen Devices} \\
                    & RX1 & RX1+2 & RX1+2+3 \\
                    \hline
                    \textbf{Single-day} & 81.00\% & 95.00\% & 100.00\% \\
                    \textbf{Multi-day} & 42.00\% & 71.00\% & 100.00\% \\
                    \hline
                \end{tabular}
                \label{tab:closed_set_mobrffi_comparison}
                \vspace{-1em}
            \end{table}

    \subsection{Open-set Known/New Classification}

        A device re-identification system must not only support the classification of known devices but also identify whether a newly captured signal arrived from a new source. This is a challenging task since the quality of received fingerprints is heavily affected by the emitter location and environmental conditions. In this section, we investigate this challenge. First, we explore whether the combination of fingerprints from multiple receivers can improve known/unknown classification accuracy. Second, we evaluate the MobRFFI framework on this task across single-day, multi-day, single-receiver, and multi-receiver modalities.

        \subsubsection{Evaluation of Single-receiver and Multi-receiver Fingerprint Distance Gaps} \label{sec:multirx_reid}

            RF device fingerprints have been shown to demonstrate consistently strong performance in closed-set scenarios \cite{hanna2022wisig, shen2022towards}, which we confirmed in the previous section. However, the open-set challenge is more complex. Here, the system must first resolve a binary classification problem -- deciding whether the newly acquired signal arrives from a known (or unknown) device. 

            One can achieve this by defining an inter-device fingerprint Euclidean distance threshold. If the new fingerprint's distance to all fingerprints in the database is above the threshold, we can label this device as new. However, ever-changing environmental conditions, including the absence of line-of-sight conditions, can hinder this decision metric. 
    
\begin{figure}[t!]
                \centering
                \begin{subfigure}[b]{0.7\columnwidth}
                    \centering
                    \includegraphics[width=\columnwidth]{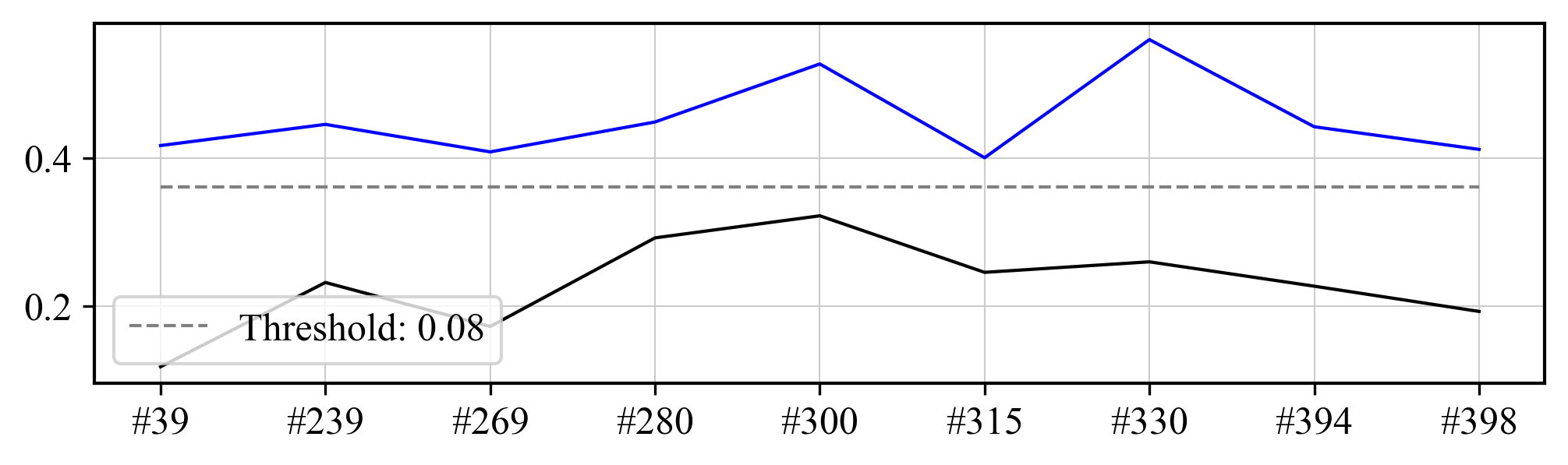}
                    \caption{One receiver.}
                    \label{fig:multi_rx_1}
                \end{subfigure}
                
                \begin{subfigure}[b]{0.7\columnwidth}
                    \centering
                    \includegraphics[width=\columnwidth]{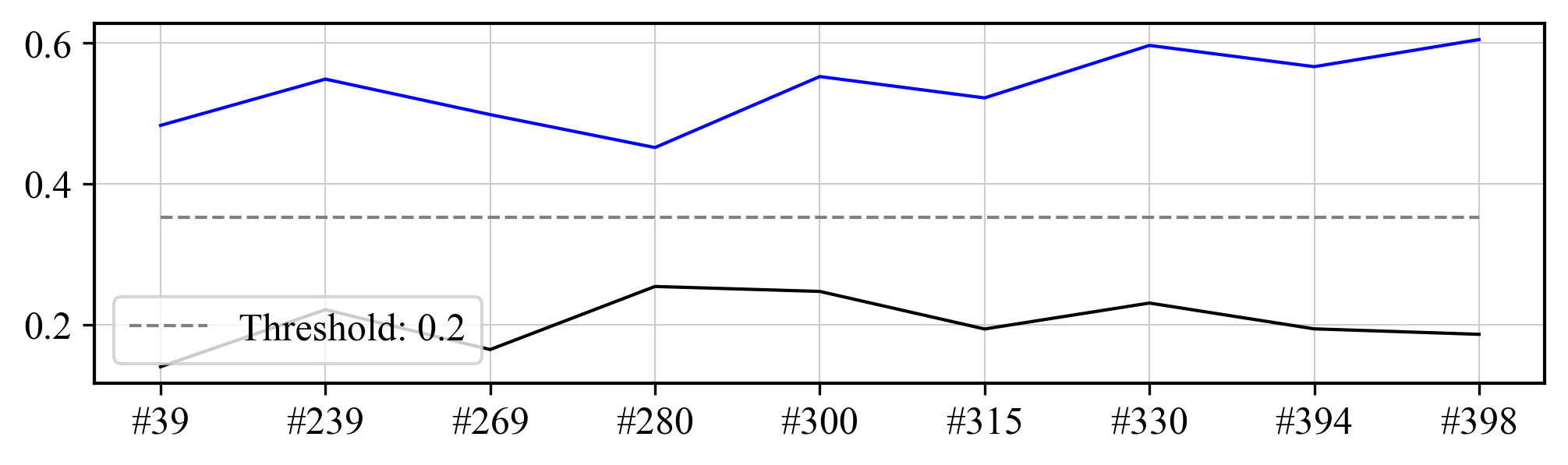}
                    \caption{Two receivers.}
                    \label{fig:multi_rx_12}
                \end{subfigure}
                
                \begin{subfigure}[b]{0.7\columnwidth}
                    \centering
                    \includegraphics[width=\columnwidth]{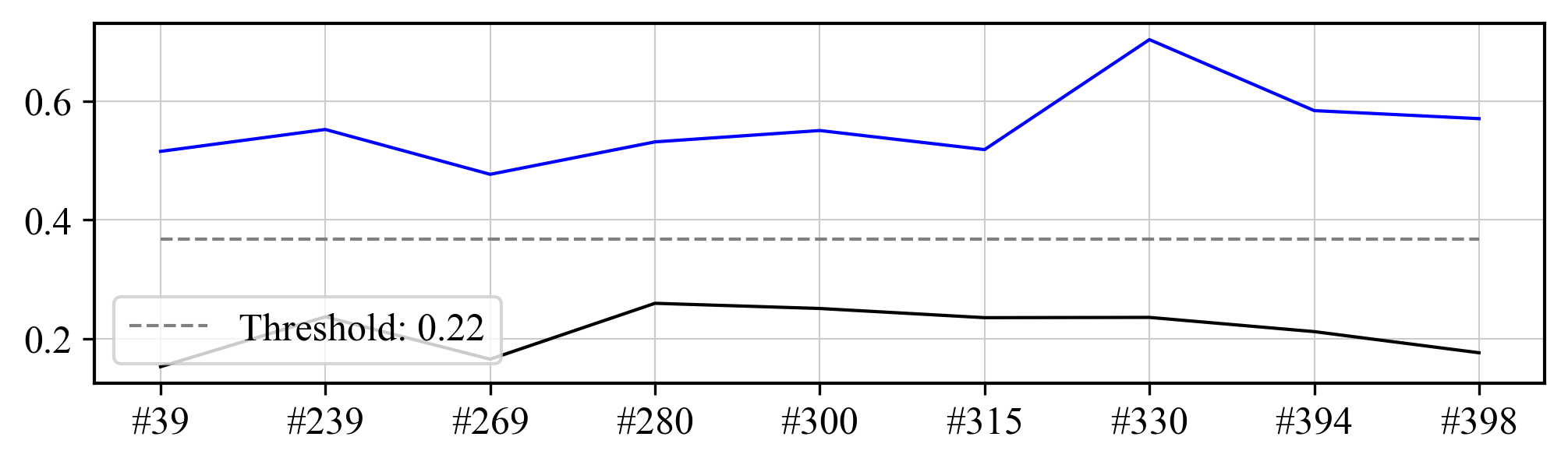}
                    \caption{Three receivers.}
                    \label{fig:multi_rx_123}
                \end{subfigure}
                
                \begin{subfigure}[b]{0.7\columnwidth}
                    \centering
                    \includegraphics[width=\columnwidth]{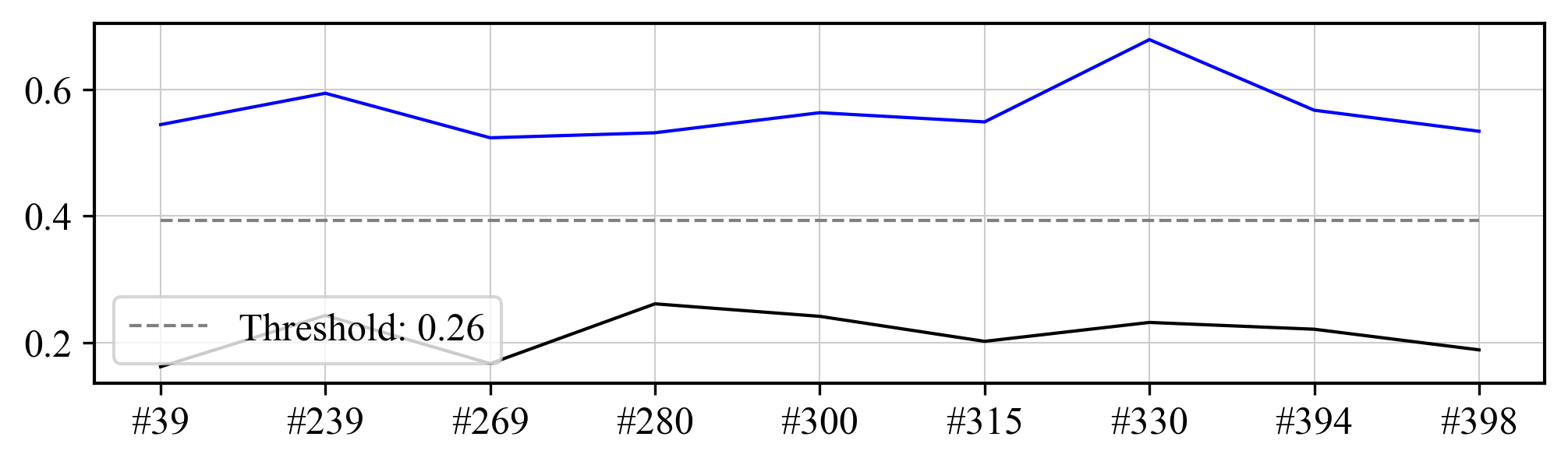}
                    \caption{Four receivers.}
                    \label{fig:multi_rx_1234}
                \end{subfigure}
                \caption{Evaluation of single-receiver and multi-receiver fingerprint distance gaps on MobRFFI dataset (\textit{August 8, 2024}).}
                \label{fig:multi_rx}
                % \Description{Evaluation of single-receiver and multi-receiver fingerprint distance gaps on MobRFFI dataset (\textit{August 8, 2024}).}
                \vspace{-2em}
            \end{figure}

            Consider the fingerprint distance evaluation illustrated in Figure \ref{fig:multi_rx_1}. Here, the horizontal axis represents the index of each transmitter in the dataset. The vertical axis represents the distance between the newly acquired fingerprint and the two closest distances in the database w.r.t. the identified transmitter. The blue and red lines represent fingerprint distances for the devices with rank 1 and 2, respectively. The dotted line represents the optimal threshold value for differentiating new and enrolled devices. In simple terms, our task is to maximize the gap between the red and blue lines, leaving as much flexibility as possible for fingerprint distance fluctuation. From this figure, we can see that the optimal threshold does not leave much space for fluctuation for devices $\#280$ and $\#398$. They could risk being misclassified and enrolled several times.

            We can mitigate this issue by combining signals simultaneously captured from multiple WiFi sensors across several physical vantage points. However, not all fingerprints are created equal. Non-LOS conditions and other effects can impact fingerprint quality. To overcome this, we introduce RSSI values from corresponding frames as a weighting factor while combining distances, as discussed earlier. The results are illustrated in Figures \ref{fig:multi_rx_12}, \ref{fig:multi_rx_123}, and \ref{fig:multi_rx_1234}.

            From the figures, we see an increase in the gap between the top 1 and top 2 device candidate fingerprint distances. Specifically, we see a $10$-fold increase in the thresholding gap -- from 0.02 to 0.21 for this example. This can be explained by the fact that certain devices, such as node \textit{$\#289$}, have limited LOS conditions. In such circumstances, fingerprints from additional receivers can balance out the impact of such conditions and provide a more stable distance metric. 
        
            % While this metric alone doesn't guarantee a sufficient gap for device thresholding, it provides a robust strategy for scaling the platform and combining signals from multiple vantage points.

        \subsubsection{Evaluation of Single-receiver and Multi-receiver Known/New Device Classification}

            We evaluate the threshold-based new/known device classifier on WiSig and MobRFFI datasets using the two models trained on the datasets for closed-set evaluation. 50 frames from an unseen set of 5 devices are used for producing enrollment fingerprints. An additional set of 5 devices and 100 frames for each are used for identification.

\begin{figure}[t!]
                % \centering
                \begin{subfigure}[t]{0.47\columnwidth}
                    \centering
                    \includegraphics[width=\linewidth]{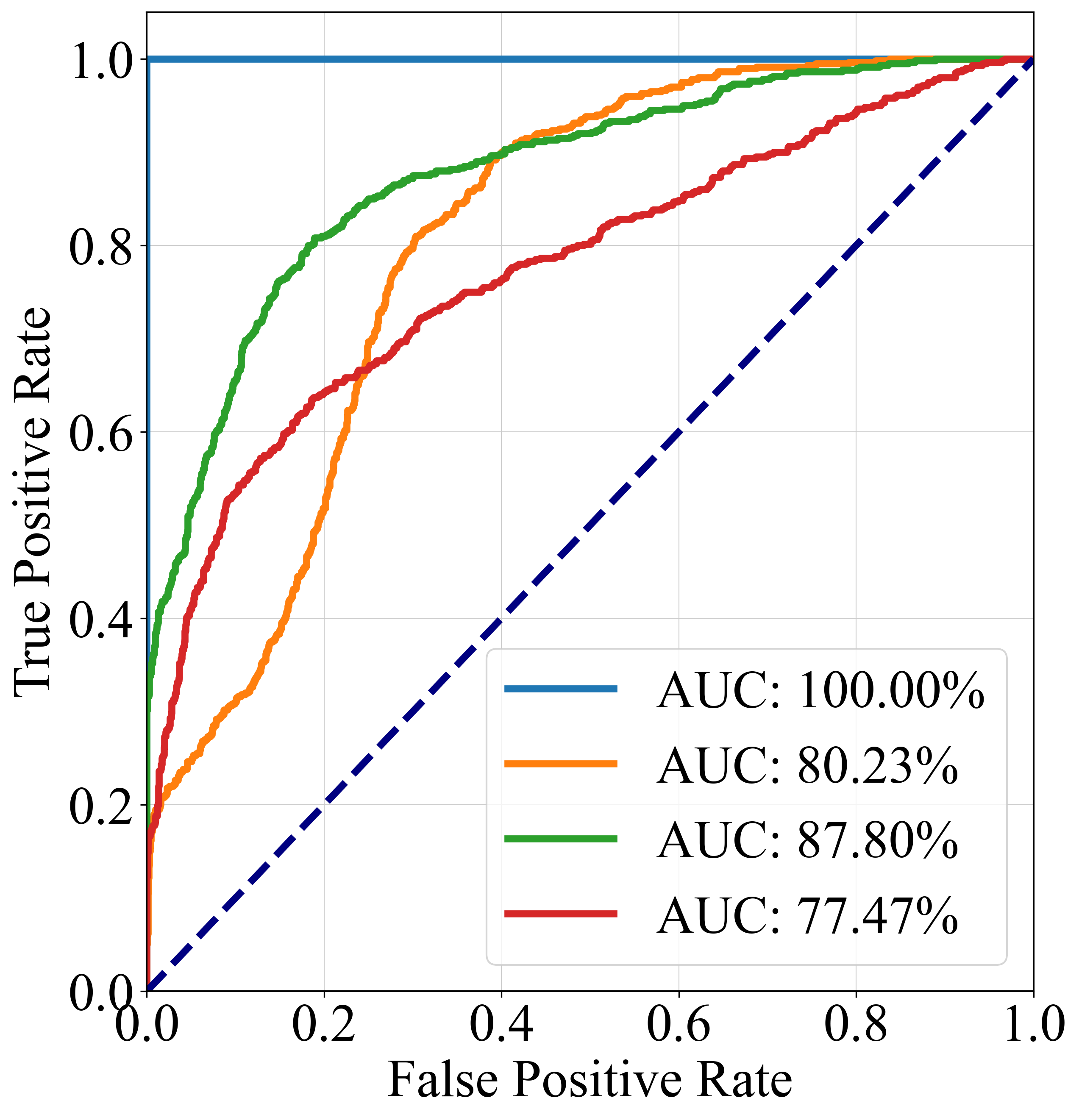}
                    \caption{WiSig dataset with single-day enroll. on Day 1, and iden. on Day 1 (blue), Day 2 (orange), Day 3 (green), and Day 4 (red).}
                    \label{fig:roc_wisig}
                \end{subfigure}
                \hspace{0.01\columnwidth}
                \begin{subfigure}[t]{0.47\columnwidth}
                    \centering
                    \includegraphics[width=\linewidth]{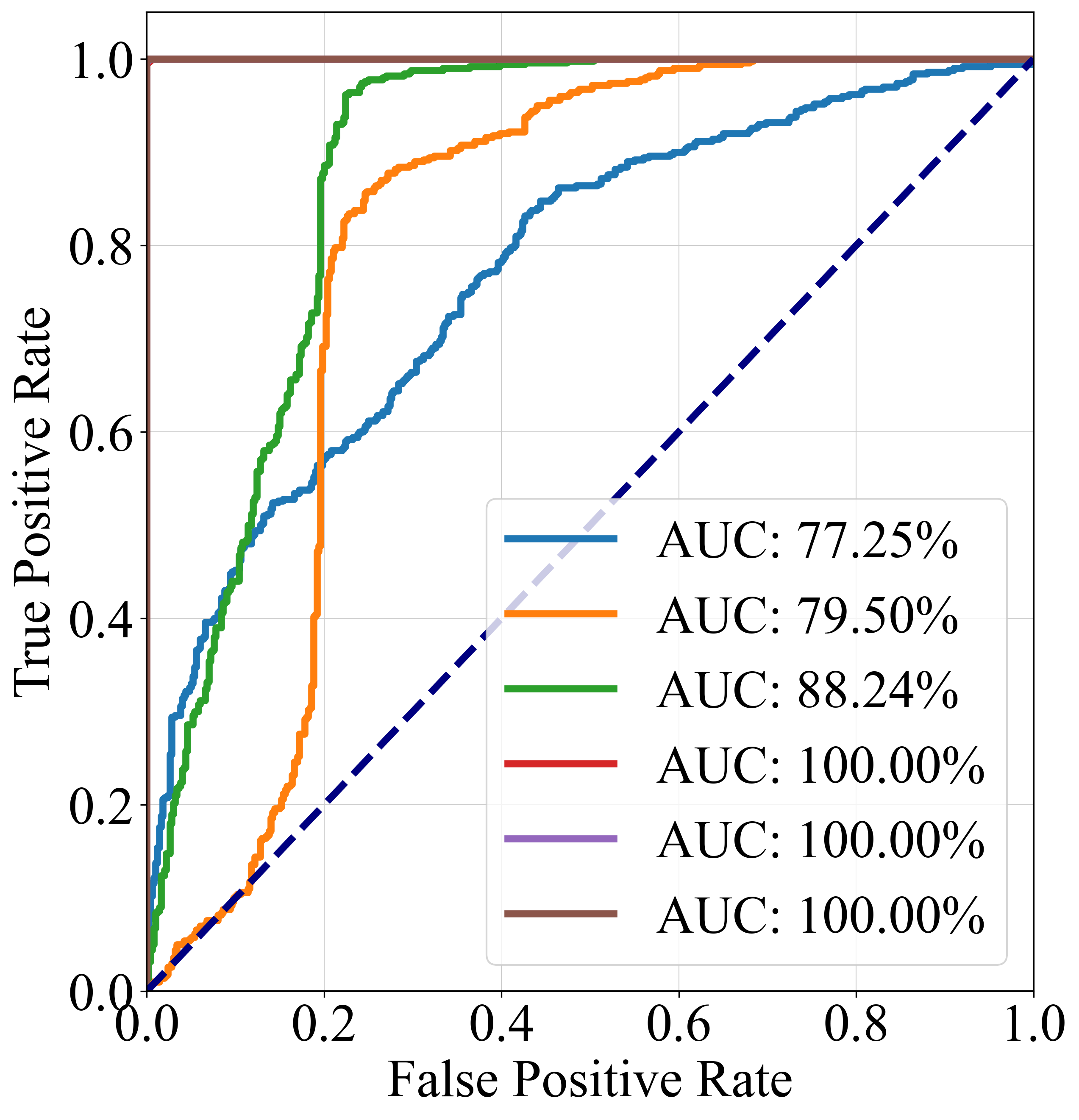}
                    \caption{MobRFFI dataset with multi-day enroll./iden. with one (blue), two (orange), and three (green) receivers, and single-day enroll./iden. (red, purple, and brown).}
                    \label{fig:roc_mobrffi}
                \end{subfigure}
                \caption{Evaluation of single-receiver and multi-receiver known/new device classification.}
                \label{fig:roc_comparison}
                % \Description{Evaluation of single-receiver and multi-receiver known/new device classification.}
                \vspace{-2em}
            \end{figure}

            The results are illustrated in Fig. \ref{fig:roc_wisig} and Fig. \ref{fig:roc_mobrffi} for the WiSig and MobRFFI datasets, respectively. Here, the horizontal axis represents the false positive classification rate, and the vertical axis represents the true positive classification rate. The diagonal line represents the random guess baseline, while the colored lines represent the receiver operating characteristic (ROC) curves and their corresponding area under the curve (AUC) values. 
            
            For the WiSig dataset, we provide the model performance results for single-day and multi-day scenarios across the three days that follow (Fig. \ref{fig:roc_wisig}). For the MobRFFI dataset, we evaluate classifier results across the two-day period and additionally evaluate performance improvements when combining fingerprints from multiple receivers (Fig. \ref{fig:roc_mobrffi}).

            For WiSig, the model delivers 100\% single-day performance, and 80\%, 88\%, and 77\% for identification on day 2, day 3, and day 4, respectively. For MobRFFI, the model achieves 100\% for single-day use and 77\%, 80\%, and 89\% for multi-day use when combining fingerprints from one, two, and three receivers, respectively.

\section{Conclusions} \label{sec:conclusions}

    In this paper, we explored an approach to re-identifying WiFi devices without relying on MAC addresses. We presented a WiFi device fingerprinting framework, MobRFFI, which implements an encoder-based deep learning architecture, producing unique WiFi fingerprints based on WiFi preamble spectrograms.

    We explored two approaches to evaluate MobRFFI. First, we adopted the state-of-the-art WiFi fingerprinting dataset, WiSig. We used it to evaluate WiFi signal pre-processing techniques and introduced a spectrogram optimization, increasing device classification accuracy by 13\%. Next, we benchmarked our model performance with these techniques across various modalities on the WiSig dataset and achieved 94\% device classification accuracy on a closed-set scenario. 

    To expand our evaluation to a more granular, multi-receiver scenario, we collected a 5.7 TB WiFi device fingerprinting dataset, MobRFFI. With the new measurements, we demonstrated that combining fingerprints from multiple receivers boosts known/unknown device classification accuracy from 81\% to 100\% on a single-day scenario and from 41\% to 100\% on a multi-day scenario.

\section*{Acknowledgment}

    This research is supported through the ERC for Smart Streetscapes, funded through NSF award EEC-2133516.

    We gratefully acknowledge the support of (1) the City of West Palm Beach; (2) Chris Roog, Executive Director of the Community Redevelopment Agency in the City of West Palm Beach; (3) Ivan Seskar, Chief Technologist at WINLAB, Rutgers University; (4) the staff of the ORBIT Testbed facility; and (5) Florida Atlantic University's \emph{I-SENSE} team.

% For local compilation (to produce *.bbl file)
% \begingroup
% \renewcommand*{\bibfont}{\footnotesize}
% \printbibliography
% \endgroup

\footnotesize
\bibliographystyle{IEEEtran}
\bibliography{references}

% For Arxiv submission
% \begingroup
% \footnotesize
% \input{main.bbl}
% \endgroup

\end{document}